\documentclass{aa}  
\usepackage{graphicx}
\usepackage[T1]{fontenc}
\usepackage{amsmath}
\usepackage{graphicx}
\usepackage{txfonts}
\usepackage{natbib} 
\bibpunct{(}{)}{;}{a}{}{,} 

\newcommand{\ve}[1]{{{\bf #1}}}

\begin{document}

\title{A magnetic thrust action on small bodies orbiting a pulsar.}


   \author{F. Mottez \inst{1}            
          \and
          J. Heyvaerts \inst{2,1}\fnmsep
          }


   \institute{LUTH, Observatoire de Paris, CNRS, Unviersit\'e Paris Diderot,
              5 place Jules Janssen, 92190 Meudon, France.\\
              \email{fabrice.mottez@obspm.fr}
         \and
             Observatoire Astronomique, Universit\'e de Strasbourg,
		11, rue de l'Universit\'e, 67000 Strasbourg, France.\\
             \email{jean.heyvaerts@astro.unistra.fr}}


 
  \abstract{}{We investigate the electromagnetic interaction of a relativistic stellar wind with small bodies in orbit around the star. 
\\ }
  {Based on our work on the theory of Alfv\'en wings to relativistic winds presented in a companion paper, we estimate the force exerted 
  by the associated current system on orbiting bodies and evaluate the resulting orbital drift.\\}
  {
  This Alfv\'enic structure is found to have no significant influence on planets or smaller bodies
  orbiting a millisecond pulsar. 
  On the timescale of millions of years, it can however affect 
  the orbit of bodies with a diameter of 100 kilometres around
  standard pulsars with a period $P \sim $1 s and a magnetic field $B \sim  10^{8}$ T.
  Kilometer-sized bodies experience drastic orbital changes on a timescale of  $10^4$ years.
  }{}

   \titlerunning{magnetic thrust}
   \authorrunning{Mottez and Heyvaerts}
   \maketitle
%

\newpage

\section{Introduction} \label{sec_intro}

Accretion discs are expected 
to form at some phase of the evolution of neutron stars in a binary system,
possibly giving birth to second generation planets. 
{{
The interaction of the disc with the pulsar's wind constrains
its extension and notably the value of its inner radius. 
The rotational power transferred to the disk by the magnetic field of a young neutron star
(propeller effect) would rapidly disrupt a disc of plasma, 
putting its inner radius beyond the light cylinder \citep{Eksi_2005}. 
Concerning dust, ablation by the pulsar's wind would extract protons and provide them 
enough kinetic energy to chase them away \citep{Jones_2007}. 
The Poynting-Robertson effect \citep{Burns_1979} 
may also act on the drift of smaller 
(and isothermal) particles such as grains and dust \citep{Cordes_2008}.  
This is consistent with infra-red observations of radio-pulsars showing that their inner 
radius is two or more orders of magnitude larger than the light cylinder radius \citep{Jones_2008}. 

On the other hand, 
small bodies, such as planetoids, asteroids or comets may also orbit pulsars   
and it is expected that some of them occasionally fall below the light cylinder.
For instance \citet{Cordes_2008} have shown that
neutral, circumpulsar debris that enters the magnetospheres of neutron stars 
could disrupt current flows  and account for some of the intermittency seen in radio pulsars. 
The neutral material of size $\sim 1$ m or more can move toward 
the star under the influence of collisions and of the Yarkovsky effect 
(a net force induced by a difference of temperature between the illuminated and warm afternoon side and the night-time face). 

In the present paper, we study the influence of a pulsar's wind on the trajectory of hypothetical larger bodies (1-100 km) through the action of the Poynting flux.
It is based on Alfv\'en wings, whose theory is described in a joint paper \citep{Mottez_2011_AWW}, hereafter (MH1).
The concept of Alfv\'en wings was initially developed by \citet{Neubauer_1980} 
to explain the interaction of Jupiter with its satellite Io. 
It says that a conducting solid body embedded in a plasma flow slower than Alfv\'en waves 
supports a system of electric currents carried by 
a stationary Alfv\'enic structure.
When the wind is relativistic albeit slower than the total Alfv\'en speed,  
the amplitude of this current can be estimated. 
In the present study, we estimate the force exerted 
by the associated current system on orbiting bodies and evaluate the resulting orbital drift.
}}
%

  \section{Alfv\'en wings}
{{  
In this section, results demonstrated in the companion paper (MH1) and useful for the present study are recalled. 

 The star is assumed to be spinning with an angular frequency  $\Omega_*$. 
 For simplicity, the theory 
 is developed for a pulsar with a magnetic dipole axis aligned with the rotation axis.
 The wind flow is assumed to be radial ($r$ direction in spherical coordinates) and the magnetic field
 in the companion's environment to be mainly azimuthal ($\phi$ direction). 
 The wind is characterized by two invariants along its flow : 
 the neutron star's magnetic flux $\Psi$, and the mass flux $f$,
 \begin{eqnarray}
f&=&\gamma_0 \rho'_0 v_0^r r^2, \\
\Psi&=&r^2 B_0^r,
\end{eqnarray}
where $v_0$ is the unperturbed wind's velocity, $B_0^r$ is the radial magnetic field,
$\gamma_0$ is the wind's Lorentz factor, and $\rho'_0$ is the proper rest mass density in the unperturbed wind's frame.
The azimuthal magnetic field is given by,
\begin{equation} 
B_0^\phi=B_0^r \frac{v_0^\phi -\Omega_* r}{v_0^r} \sim -\frac{B_0^r \Omega_* r}{c},
\end{equation}
the approximation in the right hand side term being relevant at large distance from the light cylinder.

%

The engine of the Alfv\'en wings is the convection electric field associated to the wind that appears in the reference frame of the star's companion. The wing currents are generated by a potential drop $U$ along the body of radius $R_P$,
  \begin{equation} \label{eq_unipolar_pour_application_numerique}
U = 2 R_P E_0 = \frac{2 R_P \Omega_* \Psi}{r},
\end{equation}
where $E_0= v_0^r B_0^\phi$, directed perpendicularly to the wind flow and to the magnetic field, is the convection electric field induced by the motion of the wind into the magnetic field.  
  This potential generates a system of currents that flow along the companion, then in space into the plasma, in a 
 direction which depends on the wind's magnetic field and the wind's velocity. The conductivity of the plasma part of this circuit has been evaluated by \citet{Neubauer_1980} for the non-relativistic motion of Io in the Jovian magnetospheric plasma. In the ultra-relativistic wind of a pulsar, it can be approximated very simply by
  \begin{equation}
\Sigma_A \sim \frac{1}{\mu_0 c}.
\end{equation}
(See Eq. (67-69) of (MH1) for more details.)


Then, adopting a simplified geometry, it is possible to estimate the total electric current.
\citet{Neubauer_1980} gives useful expressions for the total current $I$ flowing along an Alfv\'en wing. 
Writing $R_P$ for the body's radius, he gets:
\begin{equation} 
\label{eq_total_current}
I = 4 \, (E_0 - E_i) \, R_P \, \Sigma_A = 4 \, \left(\, \frac{\Omega_* \Psi}{r} - E_i \, \right)\,  R_P\, \Sigma_A
\end{equation}
The electric field $E_i$, set along
the body, is caused by its ionosphere or surface internal resistance.
The Joule dissipation is maximum when $E_i = E_0/2$. In our estimations, we shall 
use Neubauer's values for $I$. 
%




\section{A magnetic thrust} \label{sec_force}

The above theory, because of the involved symmetries, describes mainly what happens in space, 
far enough from the body. At closer distances, the plasma suffers compressive motions
and 
compressive MHD waves certainly have a non-negligible influence on the system. These waves
propagate quasi-isotropically. Their amplitude decrases as the inverse of the distance to the body 
and they contribute to deflect the wind around it. Nevetheless, 
without entering into these consideration, we can still make a few inferences based on the theory of the Alfv\'en wings, as 
presented in the previous section. As \citet{Neubauer_1980}, we can assume 
that the current associated to the wing is closed in the vicinity of the body (see Fig. \ref{fig_inducteur_unipolaire}), 
through its surface or its ionosphere. We can estimate (roughly) what force the wind exerts on it.

The two Alfv\'en wings carry a current that, in the two branches flowing along the body (perpendicular to the plane of Fig \ref{fig_inducteur_unipolaire}), 
generates a force density $\ve{j} \times \ve{B}$. 
The two current systems flowing on each side of the body exert this force density in the same radial direction. 
We may expect the body to orbit near the equatorial plane of the pulsar. In this plane, at such a distance, the magnetic field
direction is almost azimuthal, being perpendicular to the wind flow velocity. 
The sign of $B_0^\phi$ depends on whether the magnetic moment of the neutron star is parallel or antiparallel 
to the rotation axis. Nevertheless,  the force density $\ve{j} \times \ve{B}$ always has the same direction as the wind velocity. 

At first order, considering Eq. (\ref{eq_unipolar_pour_application_numerique}), 
$E_0=\Omega_* \Psi /r$ and the force is expressed explicitely as a function of the distance $r$ from the pulsar to the body as 
\begin{equation} \label{eq_force_magnetique}
F=2 R_P I B^\phi =8 \, \left(\frac{\Omega_* \Psi}{r} - E_i\right) \, R_P^2 \, \frac{\Omega_* \Psi}{\mu_0  c^2 r} \, .
\end{equation}
The power $\dot E_{J}$ dissipated by Joule effect along the ionosphere or in the body
is maximized when the internal load matches the external one, that is, 
still according to \citet{Neubauer_1980}, when $E_i=E_0/2$.
In that case the force is 
\begin{equation} \label{eq_force_magnetique_max}
F=\frac{4}{\mu_0  c^2}  \frac{R_P^2 \Omega_*^2 \Psi^2}{ r^2}
\end{equation}
On the night side of the body, this force tends to wipe out the ionosphere (if there is one), but on the day side, 
on the contrary, it pushes the ionosphere towards it. 
The dynamics of this system is probably quite complex, but we can retain that there is a force pushing the body
and/or its atmosphere away from its star. Maybe the atmosphere has been completely wiped out, 
and the day side of the body is ionised by the flux of X rays coming from the neutron star. 
Then, the current may flow along the dayside of the body's crust, directly pushing it away.
Actually, it remains that, when the field is azimuthal and the dissipation is maximal,
the force is radial and proportional to $r^{-2}$ (Eq. (\ref{eq_force_magnetique_max})). 
Therefore, it acts the same way as the gravitational force, and cannot have a secular influence on the orbit. 
When $E_i$ does not vary with the distance $r$ of the body as $E_0$ does, according to Eq. (\ref{eq_force_magnetique}), a fraction of the force 
is not of a Keplerian nature. For an azimuthal field, 
this force is however still central and causes nothing more than a periastron precession. 
Since we are mainly interested in the evolution of the semi-major axis and of the eccentricity, we don't consider this case any further in this paper. 

But the unperturbed magnetic field also has a small radial component
and the force is therefore not exactly central. 
Our estimate that $B^\phi \sim B^r \Omega_* r/c$ shows that the 
small angle  between the magnetic field and the azimuthal direction is:
\begin{equation} \label{eq_angle_champ_magnetique}
\delta = c/\Omega_* r \, .
\end{equation}
In the case of PSR 1257+12, at 1 AU, $\delta =  2\times 10^{-6}$.
In the general context of a vacuum dipole wave, or a pulsar wind, its sign does not vary, 
and its amplitude decreases gently with the distance. 
As the force density $\ve{j} \times \ve{B}$ is perpendicular to the magnetic field, 
the force is not strictly radial when the field is not strictly azimuthal. In spite of the small value of this angle, 
this azimuthal force component acts constantly in the same direction. Therefore, this force can work.

The tangential component of the force always has
the same direction as the rotation of the neutron star. Therefore, if the planet's orbital angular momentum
 and the star's rotational spin are parallel (in the same direction), the $\ve{j} \times \ve{B}$ force 
contributes to its acceleration, and therefore it  increases its semi-major axis and its eccentricity. 
The force modulus increases at smaller distance and the angle $\delta$ also becomes larger.
These two effects cause the tangential force to become stronger at closer distances from the star. 

From Eqs. (\ref{eq_force_magnetique_max}) and (\ref{eq_angle_champ_magnetique}), a rough estimate of the tangential force is 
\begin{equation} \label{eq_force_tangente}
F_t= F \delta =  \frac{4}{\mu_0  c}  \frac{R_P^2 \Psi^2 \Omega_*}{ r^3}.
\end{equation}
Let $v_{orb} \sim (G M_*/r)^{1/2}$ be the orbital velocity. The power associated to the work of $F_t$ is
\begin{equation} \label{eq_travail_force_tangente}
\dot W_t= F_t v_{orb} \sim  \frac{4 G^{1/2}}{\mu_0  c}  \frac{M_*^{1/2} R_P^2 \Psi^2 \Omega_*}{ r^{7/2}}.
\end{equation}

Four planets have been discovered around two pulsars. Three of them orbit PSR 1277+12, with periods of the order of a few weeks. Their mass is comparable to the mass of the Earth (more details are given in (MH1)). Planets around pulsars are expected to have been captured by the neutron star or 
to have orbited 
its progenitor before the supernova explosion.
A capture would provide an initially large orbital eccentricity. 
Similarly, a body surviving a supernova explosion 
should be left after the event with a large eccentricity.
However, the eccentricities of the orbits of planets measured around pulsars are very small
(see Table (2) in (MH1)). Could the tangential component 
of the $\ve{j} \times \ve{B}$ force associated to the planets'Alfv\'en wings be an explanation ? 

Let us notice that a captured planet orbiting in the opposite direction to the star's rotation
experiences a tangential component of the $\ve{j} \times \ve{B}$ force that tends to
slow it down and/or to reduce its eccentricity.

This force may also be exerted on smaller bodies, such as comets or asteroids. Here again, it is interesting to know how
the orbits of such bodies would be influenced by their Alfv\'en wings. 
This may be of importance for  second generation planets, which
form (or not) after the supernova explosion
from solid debris in the fall back accretion disc.

\section{Influence of the magnetic thrust on the orbit}
Let us now write down the equations of motion of an isolated body, orbiting a neutron star, under the action 
of the (Newtonian) gravitational force and the magnetic thrust.
The magnetic thrust is decomposed into its radial component (Eq. (\ref{eq_force_magnetique_max}))
and orthoradial component. We roughly
assume $E_i$ to be a fraction of $E_0$.
The  orthoradial component is given by Eq. (\ref{eq_force_tangente}). 
The acceleration then is:
\begin{equation}  \label{FtetD} 
\frac{F_r}{M_P}=\frac{C}{r^2} \quad \mbox{ and } \quad \frac{F_t}{M_P} = \frac{D}{r^3},
\end{equation}
{{where $C$ and $D$ are constant factors,}}
\begin{eqnarray}
C &=&\frac{4 R_P^2 \Omega_*^2 \Psi^2}{\mu_0 c^2 M_P}, \\
D &=& \frac{4 R_P^2 \Omega_* \Psi^2}{\mu_0 c M_P}.
\end{eqnarray}
The radial force does not modify the body'orbit, which remains Keplerian, 
the star's mass $M_*$ being
replaced by the slightly lower mass $M$, 
\begin{equation}
M = M_* - \frac{C}{G}\, .
\label{Mallegee}
\end{equation}
The equations of motion then become
\begin{eqnarray}
&& {\ddot{r}} - r {\dot{\phi}}^2 = - \frac{GM}{r^2} 
\label{equationradiale1} \\
&&\frac{d \, (r^2 {\dot{\phi}}) }{dt} = \frac{D}{r^2}
\label{eqmomentang1}
\end{eqnarray}
These equations are those of a Keplerian motion, with a small correction induced by the right-hand side term of Eq. (\ref{eqmomentang1}). Therefore, we can consider that this motion is Keplerian in first approximation, and that
the orbital elements $a$ and $e$ evolve very slowly. They are quasi-constant over an orbit. 
It is possible to compute their  slow average variations  over an orbital period, $< da/dt>$ and $<de/dt>$.
One first needs to compute the instantaneous drifts $d a/d\phi$ and $d e /d\phi$, and the variations
$\Delta a$ and $\Delta e$ over an orbit through an integration over $\phi$ from 0 to $2\pi$, 
considering the values of $a$ and $e$ to be constant in this sum and the motion to be purely Keplerian.
The average values $< da/dt>$ 
and $<de/dt>$ are the variations $\Delta a$ et $\Delta e$ divided by the orbital period. 
The orbit of a Keplerian motion is represented by the equation
\begin{equation}
r = \frac{a \, (1 - e^2)  }{1 + e\, \cos \phi}
\label{eqtrajectoire}
\end{equation}
The correction (proportional to $D$) is equivalent to a force that is tangential to the orbit. 
The induced variations of $a$ and $e$ in this rather standard problem are \citep{Milani_1987}:
\begin{eqnarray}
\frac{d a}{d\phi} &=& \frac{2 D}{GM} \ \frac{(1 + e\, \cos \phi)^2  }{(1 - e^2)^{2} }
\label{dasurdphi}
\\
\frac{ d e  }{d\phi} &=& \frac{D}{GMa} { \frac{e(1 + \cos^2 \phi) + 2 \cos \phi }{(1- e^2)}}
\label{desurdphi}
\end{eqnarray}
In one orbit, $a$ changes by
\begin{equation}
\Delta a = \int_0^{2\pi} \frac{d \, a}{d\phi} \ d\phi= \frac{4 \pi D}{GM} \ \frac{1 + e^2/2}{(1 - e^2)^2}
\end{equation}
The change of the eccentricity over one orbit is
\begin{equation}
\Delta e= \int_0^{2\pi} \frac{d \, e}{d\phi} \ d\phi= \frac{3 \pi D}{GMa} \ \,  \frac{e}{(1- e^2)},
\end{equation}
The average changes of $a$ and  $e$ over an orbital period are
\begin{eqnarray} \label{eq_delta_a}
< \frac{da}{dt}> &=& \frac{\Delta a }{P} =  2 a \, \frac{D}{\sqrt{GM a^5}} \ \, \left(\frac{2 + e^2}{2(1 - e^2)^{2}}\right)
\\ \label{eq_delta_e}
< \frac{de}{dt}> &=& \frac{\Delta e}{P} =  \frac{3}{2} \, \frac{D}{\sqrt{GM a^5}} \ \, \frac{e}{(1- e^2)}
\end{eqnarray}
These variation rates both have the same sign as $D$.
We can see that for a prograde orbit, $D>0$, $a$ and $e$ increase, the orbit becoming more eccentric and distant. 
Therefore, the Alfv\'en wing thrust tends to chase the body away from the star. A retrograd orbit 
evolves toward a circular shape with a decreasing semi-major axis.

We now present a few numerical applications of Eqs. (\ref{eq_delta_a}) and (\ref{eq_delta_e}). The 
basic numerical data about the pulsars and their companions can be found in Tables 1 and 2 of the joint paper (MH1).
In table \ref{table_application_planetes} of the present paper, we have written the corresponding yearly variations of $a$ and $e$ 
for the four know pulsar's planets, and for hypothetical pulsar's companions.
For large planets, the influence of the Alfv\'en wing on the orbit is negligible; 
the Alfv\'en wing could not explain why the orbits of the planets around PSR 1257+12 and PSR 1620-26 have a small eccentricity.
The orbit of small bodies (1 and 100 km) orbiting a 10ms pulsar is also not significantly altered by the Alfv\'en wing. 
On the contrary, the trajectory of a 100km body orbiting a 1s pulsar, because of a much larger ambient magnetic field, can be significantly modified on a time scale of millions of years. The effect is significant in only 10 000 years for a 1km sized asteroid. 
The orbits of asteroids orbiting in the sense of the neutron star's rotation spin 
increase in size and become more and more eccentric.
Objects in counter-rotation (anti-parallel orbital angular momentum and rotation spin) would
be quickly precipitated onto the neutron star. We can then expect that a young pulsar, even isolated (with no accretion disc), can stimulate the in-fall of small bodies.

{{
In order to get a more precise idea about the Eqs. (\ref{eq_delta_a},\ref{eq_delta_e}), we have solved them numerically, using a fourth order Runge-Kutta algorithm, for a small set of initial orbits. We have solved the equation over a time span small enough to consider that the pulsar parameters  that determine the parameter $D$ are constant. Figure \ref{fig_time_aaa_1km_plus} shows the evolution of the semi-major axis of a 1 km size body, initially at a distance of $0.16$ AU, for various values of the initial eccentricity. We have chosen $D/\sqrt{GM}=+10^{14} \mbox{m}^{5/2} \cdot \mbox{s}^{-1}$ that corresponds to the value given in Table \ref{table_application_planetes}. The sign $+$ means that the body orbits in the sense of the pulsar' spin. We can see that, in accordance to what was said above, the semi-major axis increases significantly in a time scale of tens of thousand years. The increase is slightly larger for an initially large eccentricity.  Figure \ref{fig_time_eee_1km_plus} shows the evolution of the corresponding eccentricities. There is no variation for an initially null eccentricity, and the larger the initial eccentricity, the larger its further increase.

Figure \ref{fig_time_aaa_1km_moins} shows the evolution of the semi-major axis for the same body in the
case of counter-rotation. In less than 6000 years, the body falls onto the star, or at least beyond the inner 
frontier between the wind and the magnetosphere (which is the limit of validity of the present calculations). 
In the inner magnetosphere, the magnetic thrust still acts onto the body (although differently), and a rapid fall onto the star
is expected.  }}

\begin{table*}
\caption{Electric potential drop, total electric current associated to the Alfv\'en wing. 
Electrical energy $\dot E_{J max}$ dissipated in the Alfv\'en wing. 
Variation per (terrestrial) year of the semi-major axis  
estimated from Eq. (\ref{eq_delta_a}). 
Variation of the eccentricity, per year, $\Delta e /year$, estimated from Eq. (\ref{eq_delta_e}).}  
\label{table_application_planetes} 
\centering 
\begin{tabular}{l r r r r  r r} 
\hline\hline 
Name & $U$ (V)& $I$ (A)& $\dot E_{J max}$ (W)& $\Delta a/year$ (m) & $\Delta e /year$ & $D/\sqrt{GM}$ (m$^{5/2} \cdot $s $^{-1}$)\\ 
\hline 
PSR 1257+12 a &  1.1 $\times 10^{12}$ & 3.0 $\times 10^{9}$ & 2.5$\times 10^{21}$ &  0.02 & 0 &3.06$\times 10^{6}$\\
PSR 1257+12 b &  3,5 $\times 10^{12}$ & 9.4 $\times 10^{9}$ & 2.5$\times 10^{22}$ &  $ 1.\times 10^{-3}$ & 3 $ \times 10^{-16}$ &5.11$\times 10^{5}$\\
PSR 1257+12 c &  2,6 $\times 10^{12}$ & 7.0 $\times 10^{9}$ & 1.4$\times 10^{22}$ &  $ 1.\times 10^{-3}$ & 2.4$ \times 10^{-16}$ &5.28$\times 10^{5}$\\
\hline 
PSR 1620-26 a &  6,0 $\times 10^{11}$ & 1.5 $\times 10^{9}$ & 7$\times 10^{20}$ &   3.$\times 10^{-6}$  & 0& 6.34$\times 10^{5}$\\
\hline 
PSR 10ms b 100 km &   2.4$\times 10^{9}$ &  6.$\times 10^{6}$ & 1.2$\times 10^{16}$ &  $ 8.\times 10^{-3}$&   6.$\times 10^{-14}$&  9.94$\times 10^{5}$\\
PSR 10ms b 1 km   &   2.4$\times 10^{7}$ &  6.$\times 10^{4}$ & 1.2$\times 10^{12}$ &  0.8 &  6.$\times 10^{-12}$&  9.94$\times 10^{7}$\\
\hline 
PSR 1 s b 100 km &  2.4  $\times 10^{11}$ &  6$\times 10^{8}$ & 1.2 $\times 10^{20}$ & 8$\times 10^{4}$ & 6.4 $\times 10^{-8}$ &  9.94$\times 10^{11}$\\
PSR 1 s b 1 km   &  2.4 $\times 10^{9}$ &  6.$\times 10^{6}$ & 1.2$\times 10^{16}$ &  8.$\times 10^{5}$  & 6.$\times 10^{-6}$  &  9.94$\times 10^{13}$\\
\hline 
\end{tabular}
\end{table*}


\section{Conclusion}
{{
A planet orbiting around a pulsar develops a system of Alfv\'en wings, caused by its interaction 
with the sub-Alfv\'enic Poynting-flux-dominated pulsar wind. A system of strong electric currents is set, 
which exerts an ortho-radial force upon the planet that can, if the magnetic to mechanical energy coupling is efficient, 
have an incidence on the planetary orbit. 

The  data in Table \ref{table_application_planetes} 
show that the orbital drift of a planet around a millisecond pulsar is negligible. In particular, this effect cannot be involved in an explanation for the very low eccentricity of the four known planets orbiting a pulsar.
But, for bodies with a diameter of a few kilometres orbiting around a $P=1$ second pulsar, 
the drifts occur  on time scales that are short in comparison with the time of evolution of an isolated pulsar.
}}
For bodies orbiting in the same direction as the star'spin  (the orbital angular momenta  being in the same direction as the star's angular momentum) this force tends to increase the semi-major axis and the eccentricity. For bodies in counter-rotation (the two angular momentum having opposite directions), this force tends to decrease the eccentricity and the semi-major axis, favouring the precipitation of the body onto the neutron star. 
 If the bodies falling into the pulsar's magnetosphere studied by \citet{Cordes_2008} fall under the influence of the Alfv\'en wings, then these object were, before their fall, in counter-rotation with the the neutron star. Rocks and asteroids in counter-rotation may come directly from  the fall back after the supernova explosion. They may as well come from recent collisions. {{As suggested by \citet{Cordes_2008} the smaller bodies (about 1 metre) falling toward the star, and evaporating in its vicinity, may cause a momentary interruption of the pulsar's radio emissions. It is also possible that the fall of a larger object on the star's crust powers transient high energy emissions, such as those observed with soft gamma-ray repeaters. This might be particularly relevant for those associated to pulsars with a "standard" magnetic field \citep{Rea_2010} or with the Crab nebula \citep{Abdo_2011}. These aspects of the question remain to be explored}}

More generally, the force caused by Alfv\'en wings may have consequences on the dynamics of fall-back accretion discs, and onto the formation of second generation planets around pulsars.

\begin{acknowledgements}
The authors thank Silvano Bonazzola (LUTH, Obs. Paris-Meudon) for interesting comments on this topics. 
This research has made use of the SIMBAD database,
operated at CDS, Strasbourg, France, and The Extrasolar Planets Encyclopaedia 
(http://exoplanet.eu/index.php), maintained by Jean Schneider at the LUTH, and the SIO, at the Observatoire de Paris, France.

\end{acknowledgements}



\begin{figure}
\resizebox{\hsize}{!}{\includegraphics{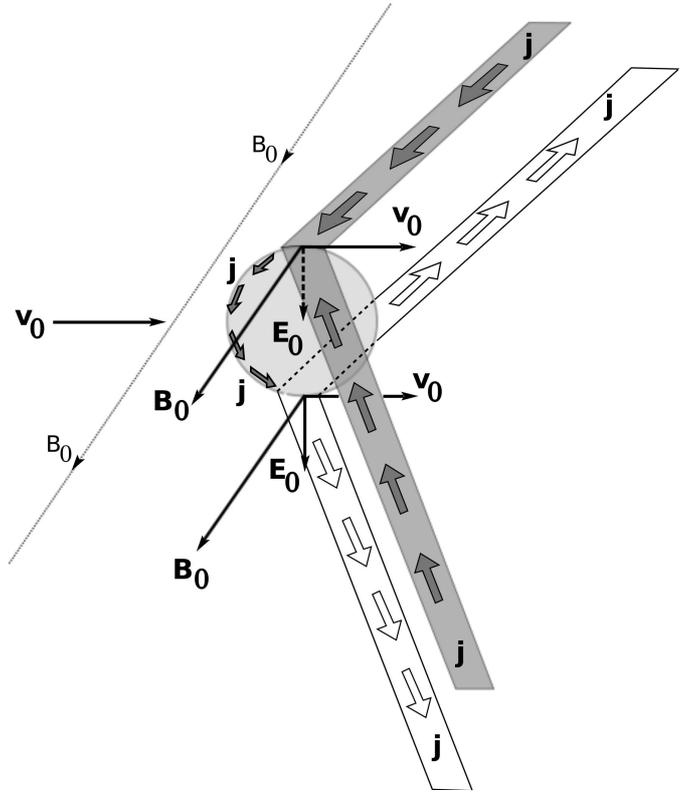}}
\caption{Schematic view of an unipolar inductor. 
The unperturbed wind's magnetic field $\ve{B}_0$ and  velocity $\ve{v}_0$ are almost, but not exactly, perpendicular. The electric field $\ve{E}_0$ created by the unipolar inductor is perpendicular to these two vectors; it induces an electric current (of density $\ve{j}$) along the body. This current then goes into the interplanetary medium, forming two structures, each of them made of an outward and an inward flow. The current density $\ve{j}$ flowing along the planet is the cause of a $\ve{j} \times \ve{B}$ force density that is the topics of the present study.}
\label{fig_inducteur_unipolaire}
\end{figure}

\begin{figure}
\resizebox{\hsize}{!}{\includegraphics{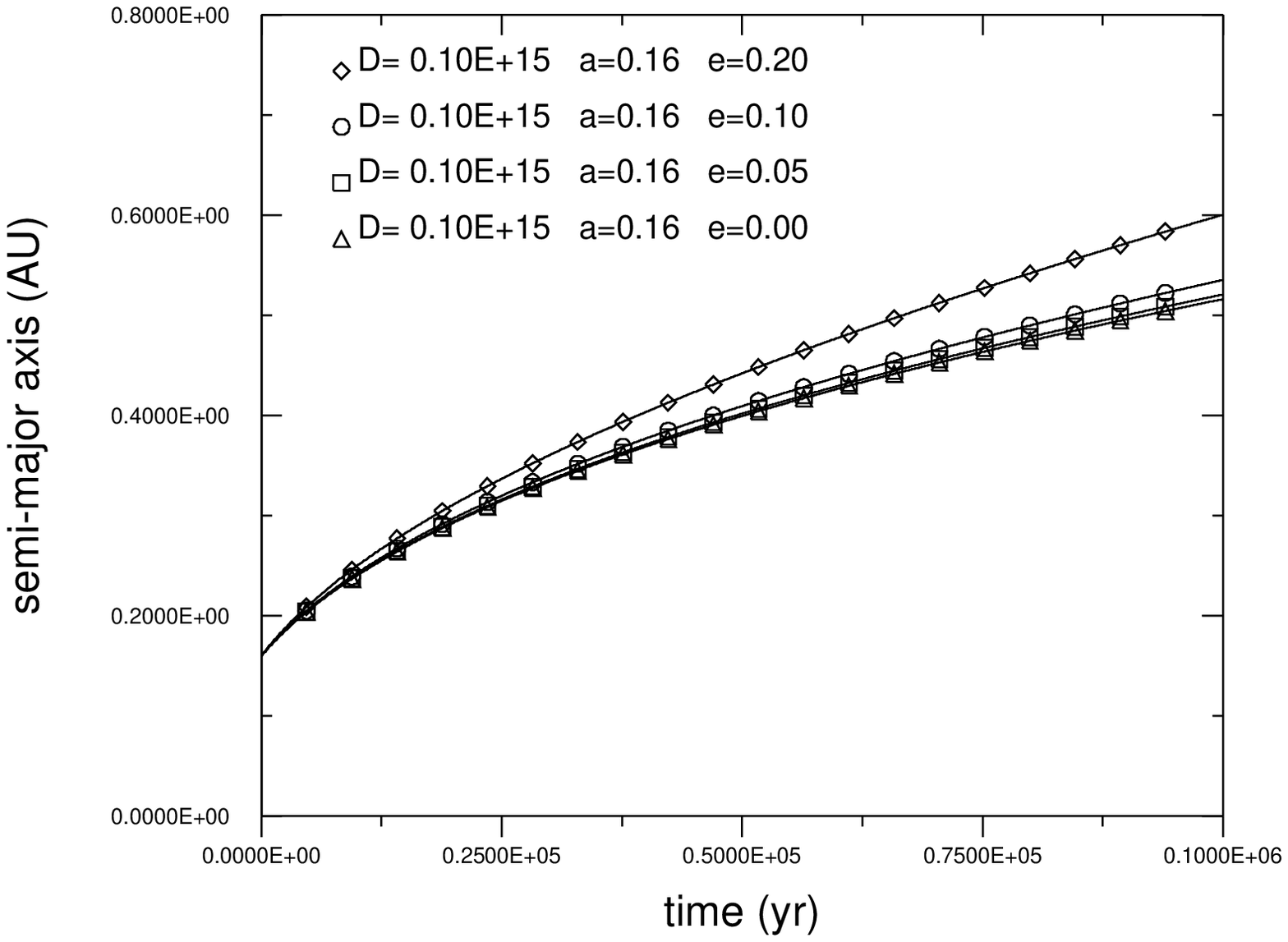}}
\caption{Evolution of the semi-major axis of a 1 km sized body as a function of time, under the influence of the magnetic thrust. The body orbits in the same direction as the pulsar' spin. The four curves are given for four different values of the initial eccentricity.}
\label{fig_time_aaa_1km_plus}
\end{figure}

\begin{figure}
\resizebox{\hsize}{!}{\includegraphics{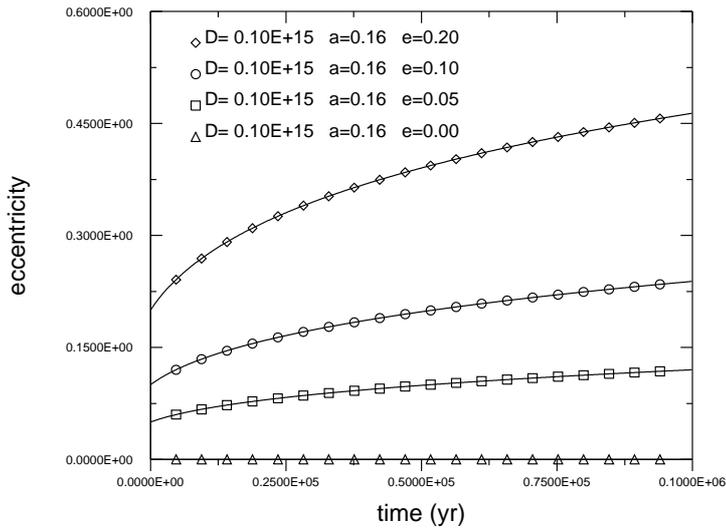}}
\caption{Evolution of the eccentricity in the same cases as in Fig. \ref{fig_time_aaa_1km_plus}.}
\label{fig_time_eee_1km_plus}
\end{figure}

\begin{figure}
\resizebox{\hsize}{!}{\includegraphics{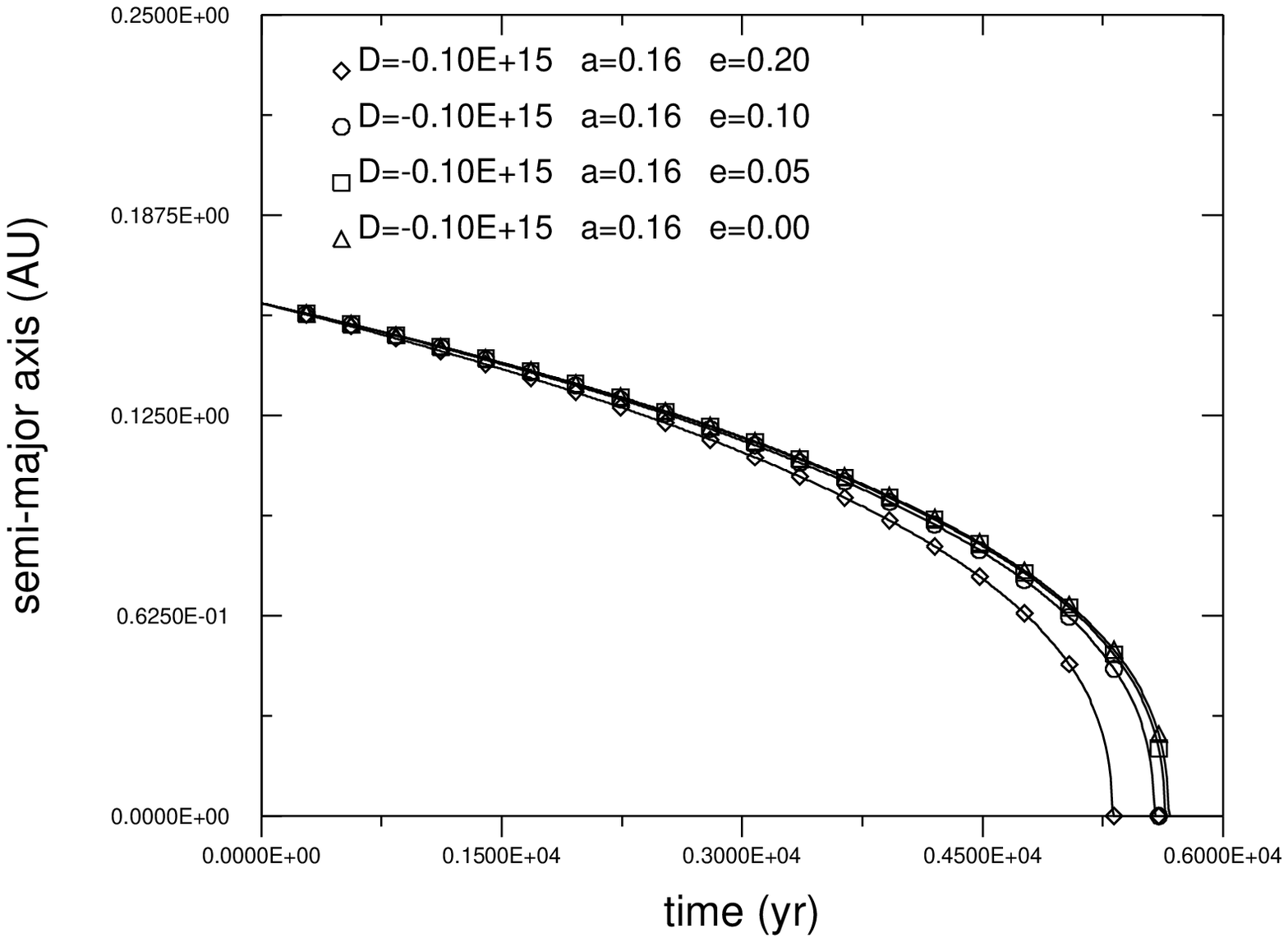}}
\caption{Evolution of the semi-major axis of a 1 km sized body as a function of time, under the influence of the magnetic thrust. The body orbits in the opposite direction to the pulsar' spin. The four curves are given for four different values of the initial eccentricity.}
\label{fig_time_aaa_1km_moins}
\end{figure}

\end{document}